\documentstyle[epsf]{mn}
\font\japit = cmti10 at 10truept
\title
     [Nonlinear Power Spectra]
{\vglue-3.0truecm
\centerline{\japit For submission to Monthly Notices}
\vglue 2.5truecm
\noindent
          Nonlinear Cosmological Power Spectra in Real and 
	  Redshift--Space 
\author
     [A. N. Taylor \& A. J. S. Hamilton ]
     {A. N. Taylor$^*$ \& A. J. S. Hamilton$^{\dag}$ \\
     $^*$Institute for Astronomy, 
     University of Edinburgh,
     Royal Observatory,
     Blackford Hill, 
     Edinburgh, 
     U.K.\\	
	$^{\dag}$ 
	JILA and Dept.\ of Astrophysical, Planetary and Atmospheric Sciences,
	Box 440, University of Colorado, Boulder CO 80309, USA\\
	ant@roe.ac.uk, ajsh@dark.colorado.edu}}

\def\bib{\parskip=0pt\par\noindent\hangindent\parindent
    \parskip =2ex plus .5ex minus .1ex}
\newcommand{\be}{\begin{equation}}
\newcommand{\ee}{\end{equation}}
\newcommand{\ba}{\begin{eqnarray}}
\newcommand{\ea}{\end{eqnarray}}

\newcommand{\et}{et al. }

\newcommand{\rgl}{\rangle}
\newcommand{\lgl}{\langle}

\newcommand{\x}{{\bmath x}}

\newcommand{\s}{{\bmath s}}
\newcommand{\q}{{\bmath q}}
\newcommand{\k}{{\bmath k}}
\newcommand{\vb}{{\bmath v}}
\newcommand{\z}{{\bmath z}}
\newcommand{\hz}{\hat{\z}}

\newcommand{\hq}{\hat{q}}

\newcommand{\xib}{{\bmath \xi}}
\newcommand{\el}{\ell}
\newcommand{\Pl}{{\cal P}_{\el}}

\newcommand{\hxb}{\hat{\x}}

\newcommand{\nn}{\nonumber \\}
\newcommand{\p}{\parallel}
	\newcommand{\n}{\perp}

\newcommand{\erf}{{\rm erf}}
\newcommand{\im}{{\rm i}}
\newcommand{\dd}{{\rm d}}
\newcommand{\lin}{{\rm L}}
\newcommand{\nonlin}{{}}
\newcommand{\Real}{{\cal R}e}

\newcommand{\K}{{\bmath K}}
\newcommand{\bzero}{{\bmath 0}}

\newcommand{\xibs}{\xib^s}
\newcommand{\psis}{\psi^s}
\newcommand{\Ps}{P^s}
\newcommand{\sk}{\tilde k}
\newcommand{\tq}{\tilde q}
\newcommand{\sP}{\tilde P}
\newcommand{\sPs}{\sP^s}
\newcommand{\sPsl}{\sPs_\el}
\begin{document}
\maketitle

\begin{abstract}



	We present an expression for the nonlinear evolution of the
cosmological power spectrum based on following Lagrangian
trajectories. This is simplified using the Zel'dovich
approximation to trace particle displacements, assuming 
Gaussian initial conditions. The model is found to
exhibit the transfer of power from large to small scales
expected in self-gravitating fields. 
Some exact solutions are found for power--law initial spectra.

We have extended this analysis into redshift--space and found 
a solution for the nonlinear, anisotropic redshift--space power
spectrum in the limit of plane--parallel redshift distortions.
The quadrupole--to--monopole ratio is calculated for the 
case of power--law initial spectra. We find that the
shape of this ratio depends on the shape of 
the initial spectrum, but when scaled to linear theory depends
only weakly
on the redshift--space distortion parameter, $\beta$.
 The point of zero--crossing 
of the quadrupole, $k_0$, is found to obey a simple scaling relation
and we calculate this scale in the Zel'dovich approximation.

This model is found to be in good agreement
with a series of $N$-body simulations
on scales down to the zero-crossing of the quadrupole, although
the wavenumber at zero-crossing is underestimated. 
These results are applied  to the quadrupole--monopole ratio
found in the merged QDOT plus 1.2 Jy {\it IRAS} redshift survey.
Using a likelihood technique we have estimated that the distortion 
parameter is constrained to be $\beta>0.5$ at the $95 \%$ level.
The local primordial spectral slope is
not well constrained, but the likelihood analysis suggests $n \approx -2$
in the translinear regime.  The zero--crossing
scale of the quadrupole is $k_0=0.5 \pm 0.1 h{\rm Mpc}^{-1}$ and from this 
we infer the amplitude of clustering is
$\sigma_8=0.7 \pm 0.05$.

We suggest that the success of this model is due to nonlinear
redshift--space effects arising from infall onto caustics and is
not dominated by virialised cluster cores.
The latter should start
to dominate on scales below the zero--crossing of the quadrupole,
where our model breaks down.

\end{abstract}


\section{Introduction}

The Newtonian analysis of linear growth of perturbations in an expanding
universe is a well understood problem (Peebles 1980; Efstathiou 1990). 
The extension of this into
the nonlinear regime has proven more difficult due to the strong
mode coupling that arises in gravitational collapse, and most
progress has been made through the use of $N$-body simulations.

Actual observations of galaxies in redshift space are further complicated
(and made more interesting) by redshift distortions,
caused by peculiar velocities of galaxies along the line-of-sight.
Again, the linear problem is relatively well understood.
Kaiser (1987) has shown that linear redshift distortions
take their simplest form when expressed in Fourier space,
at least if structure is far from the observer
so that the distortions are essentially plane-parallel.
Here the redshift--space Fourier modes are related to the real--space ones by
\be
\label{Kaiser}
	\delta^s(\k) = (1+\beta \mu_\k^2) \delta(\k)
\ee
where a superscript $s$ denotes a  
redshift--space quantity,
$\mu_\k$ is the cosine of the angle between the wavevector $\k$
and the line of sight,
and $\beta$, the redshift distortion parameter,
is the dimensionless growth rate of growing modes in linear theory,
which is related to the cosmological density $\Omega$ by
(Peebles 1980)
\be
\label{beta}
	\beta \equiv {\Omega^{0.6} \over b}
\ee
in the standard pressureless Friedmann cosmology
with mass-to-light bias $b$.
It is through measuring the distortion parameter $\beta$ that 
one hopes to measure the cosmological density parameter, $\Omega$.

Kaiser's formula (\ref{Kaiser})
is valid only in the linear limit,
and for plane-parallel distortions,
neither of which approximations is well satisfied in reality.
Several authors have now addressed the problem of generalising Kaiser's formula
to the case of radial distortions,
while retaining the assumption of linearity
(Fisher, Scharf \& Lahav 1994; Heavens \&
Taylor 1995; Fisher et al. 1995; Zaroubi \& Hoffman 1994; Ballinger,
Heavens \& Taylor 1995; Hamilton \& Culhane 1996).

The main purpose of the present paper is to extend the analysis of
redshift distortions into the nonlinear regime,
retaining the plane-parallel approximation for simplicity.
Our approach to the problem is
motivated by the consideration that
the density in redshift space may appear highly nonlinear
even when the density in real space is only mildly nonlinear.
For example, a region which in real space is just turning around,
a mildly nonlinear condition,
appears in redshift space as a caustic, a surface of infinite density,
which is thoroughly nonlinear.
This leads us firstly to work in Lagrangian space (Section \ref{Lagsec}),
and secondly to adopt the Zel'dovich (1970) approximation
(Sections \ref{Zeldsec} \& \ref{Redsec}).
The Zel'dovich approximation is in effect linear theory expressed in
Lagrangian space,
inasmuch as it approximates the trajectories of particles as
straight lines with (comoving) displacements growing 
according to linear theory.

Our approach follows that of Taylor (1993),
who studied the nonlinear evolution of the power spectrum.
Comparable approaches have been used by Bond \& Couchman (1987)
to evolve the galaxy angular correlation function,
by Mann, Heavens \& Peacock (1993) to evolve the real--space 
correlation function of clusters,
and by Schneider \& Bartelmann (1995)
to evolve the real-space (unredshifted) power spectrum.
Our approach differs from that of Hivon \et (1995), who applied a 
perturbation expansion in Lagrangian space to second order to
calculate the redshift--space skewness.

For simplicity,
we assume throughout this paper that the density field is unbiased,
$b = 1$.
Generally,
the effect of  evolution, if continuity is assumed
(which is a fundamental assumption of this paper),
is to tend to drive the bias factor towards unity.
Continuity implies that the ratio of galaxy to matter density
$(1+\delta)/(1+\delta_{\rm M})$
remains constant in Lagrangian elements,
and if (somehow) a linear bias
$\delta = b \delta_{\rm M}$
is established at some early time when $\delta$ and $\delta_{\rm M}$
are both small,
then the ratio of galaxy to matter density must be close to unity,
$(1+b\delta_{\rm M})/(1+\delta_{\rm M}) \approx 1$.
It follows that the bias will be close to unity
at later, nonlinear epochs when $\delta_{\rm M}$ is no longer small.
Conversely, if bias is in fact important at the present, nonlinear epoch
(as may well be the case),
then it must be that the assumption of continuity
must break down in the not too distant past.
Indeed, it may be that continuity is violated on an ongoing basis.
In the present paper we choose to ignore this thorny problem,
and simply assume an unbiased density field.

We begin in Section \ref{Realsec} by deriving equations
which relate the Lagrangian and Eulerian descriptions of density,
and we calculate a general expression for the evolution of the power spectrum.
We invoke the Zel'dovich approximation and Gaussian
initial conditions and consider some of the general
features of the resulting nonlinear power spectrum. Some exact 
solutions for initially power--law spectra are derived.
  In Section \ref{Redsec} 
we calculate the power spectrum in redshift--space,
again in the Zel'dovich approximation,
and obtain a number of analytic and numerical results
for the observationally interesting ratio of quadrupole-to-monopole power.
In Section \ref{Obssec}, we compare the predictions of the Zel'dovich
approximation with $N$-body simulations,
and we apply our findings to analyse the quadrupole distortion measured
in the QDOT plus 1.2 Jy redshift survey.
We summarise our conclusions in Section \ref{Sumsec}.

\section{Real--space power spectra}
\label{Realsec}
\subsection{Clustering in the Lagrangian Frame}
\label{Lagsec}

 In the Lagrangian theory of fluid mechanics,
the central variable is the  displacement vector field,
rather than the density field of Eulerian space. This
displacement field, $\xib(\q,t)$, maps particles from their initial Lagrangian
coordinate, $\q$, to the Eulerian--space coordinate, $\x$, via the 
relation\footnote{
Throughout this paper we take the coordinates $\x$ and $\q$
to be comoving coordinates,
defined relative to the general Hubble expansion of the Universe.}
\be
   	\x(\q ,t) = \q + \xib(\q,t)   \label{e:1}
\ee
where $\xib(\q,t)$ is the integral of the
velocity field along the world--line of the particle:
\be
        \xib(\q,t) = \int^t dt' \, \vb[\x(\q,t'),t']\ .       \label{e:2}
\ee
The advantage of the Lagrangian formalism for gravitational collapse
was first pointed out in the seminal work by 
Zel'dovich (1970). Because the coordinate system of Lagrangian space is
nonlinear, moving with the particles themselves, this approach takes the
analysis of perturbation growth of the density field 
into the nonlinear regime while conserving
mass--density.

However, the most useful
 observable quantity in cosmology is the Eulerian density field,
$\rho(\x)$,
inferred either from variations in the galaxy distribution, or indirectly
from distortions in the Hubble flow, gravitational lensing, 
or from the microwave background. 
%
Continuity, along with the assumption of uniform initial density,
implies that the relation between the overdensity
$\delta(\x) \equiv [\rho(\x)-\bar\rho]/\bar\rho$
in Eulerian space and the Lagrangian displacement field $\xib(\q)$ is
\be
	\delta(\x) =
	\int \dd^3\!q \, \delta_D[\x-\q-\xib(\q)]
	\ - \ 1
	\ , \label{e:6}
\ee
where $\delta_D(\x)$ is the Dirac delta function.
%
The Fourier transform
\be
\label{dkx}
	\delta(\k) = \int \dd^3\!x \, e^{\im\k.\x} \delta (\x)
\ee
of the overdensity $\delta (\x)$, equation (\ref{e:6}), is
\be
	\delta(\k) = \int \dd^3\!q \, e^{\im \k.\q} (e^{\im\k.\xib} -1)
	\ ,
\ee
which relates
the Fourier space Eulerian density field $\delta(\k)$
to the Lagrangian displacement field $\xib(\q)$.

The correlation function of Fourier modes defines the power spectrum $P(k)$:
\be
	 \lgl \delta(\k) \delta^*(\k') \rgl  =
	(2 \pi)^3 P(k)   \delta_D(\k-\k')\ ,
\ee
where the angular brackets denote ensemble averaging 
and the Dirac delta-function arises from translational invariance.
Transforming to differential and
centre of mass coordinates, we find that as a result of
translational invariance only the differential terms
survive. We can then express the power spectrum by the integral equation
\be
\label{pk}
	P(\k) = \int \dd^3\!q \, e^{\im\k.\q}
	(\lgl e^{\im\k.\Delta\xib} \rgl -1)
	\ ,
\ee
where the term $\lgl e^{\im\k.\Delta\xib} \rgl $ is identified as the
generating function of the differential displacement vector field,
$\Delta \xib ( \q )$, of points separated by distance
$\q = \q_1 - \q_2$:
\be
\label{dxi}
	\Delta\xib(\q) = \xib(\q_1) - \xib(\q_2)
	\ .
\ee
Equation (\ref{pk}) for the power spectrum is valid at all times,
not merely in the linear regime.
 However, in order to solve this completely one needs to have an expression
for the displacement generating function and its evolution.
In practice we shall have to approximate this. In the next Section
we use equation (\ref{pk}) to calculate
the evolution of the power spectrum in the Zel'dovich approximation,
under the assumption of random Gaussian initial conditions.

\subsection{Evolution of the power spectrum in the 
Zel'dovich approximation}
\label{Zeldsec}


A careful analysis of the Zel'dovich approximation has been
given by Hui \& Bertschinger (1996), who consider it in the context
of local theories of gravity. For our purposes we only need the
results that the particle displacement field scales according to
linear theory:
\ba
\label{xt}
	\xib(\q,t) &=& D(t) \xib_\lin(\q) \nn 
	&=& {\im D(t) \over (2\pi)^3} \int \dd^3\!k \,
	{\k \over k^2} \delta_\lin(\k) e^{-\im\k.\q}
\ea
where $\xib_\lin(\q)$ is the linear displacement field,
and $\delta_\lin(\k)$ the linear density field,
defined at some suitably early time.
This approximation is then a first order theory in the
displacement vector field, extrapolated to arbitrary later times.
By construction the Lagrangian coordinate system preserves mass, but in
the Zel'dovich approximation
at the expense of not satisfying the Euler equation. In the Lagrangian frame 
the density field is usually evolved along fluid streamlines,
 $\delta[\q(\x),\tau]$, but equation (\ref{e:6}) allows one to
calculate the density in Eulerian coordinates.

The assumption of a Gaussian initial fluctuation field implies
that the initial displacement field, $\xib_\lin$, was Gaussian.
The Zel'dovich approximation (\ref{xt}) further implies that
the displacement field $\xib(\q)$ remains Gaussian for all time.
Equation (\ref{pk}) for the power spectrum involves
the Lagrangian generating function of the differential displacement field,
which for a Gaussian vector field is
\ba
	\lgl e^{\im\k.\Delta\xib} \rgl &=&
	\exp (- k_i k_j \lgl \Delta\xib_i \Delta\xib_j \rgl /2 ) \nn
\label{gf}
	&=& \exp (- k_i k_j [ \psi_{ij}(\bzero) - \psi_{ij}(\q) ])
\ea
where
$\psi_{ij}(\q) \equiv \lgl \xib_i(\q_1) \xib_j(\q_2) \rgl$
is the displacement correlation function.
In the Zel'dovich approximation,
the displacement correlation function $\psi_{ij}$
grows as the square of the growth factor $D(t)$
in the nonlinear as well as linear regimes:
\be
\label{psi}
	\psi_{ij}(\q,t) = [D(t)]^2 \psi_{\lin ij}(\q)
	\ .
\ee
The linear correlation function $\psi_{\lin ij}(\q)$ of displacements
is related to the linear power spectrum $P_\lin(\k)$ by
\ba
	\psi_{\lin ij}(\q) &\equiv&
	\lgl \xib_{\lin i}(\q_1) \xib_{\lin j}(\q_2) \rgl \nn
\label{psilin}
	&=& {1 \over (2\pi)^3} \int \dd^3\!k \,
	{k_i k_j \over k^4} P_\lin(\k) e^{-\im\k.\q}
\ea
which at zero separation, $\q = \bzero$, is
\be
\label{psilin0}
	\psi_{\lin ij}(\bzero) = \lgl \xi^2 \rgl \delta_{ij}
	= {\delta_{ij} \over 3 (2\pi)^3} \int \dd^3\!k \, k^{-2} P_\lin(\k)
\ee
where $\lgl \xi^2 \rgl$ is the one-dimensional dispersion of displacements.

Inserting the Gaussian generating function (\ref{gf}) into
equation (\ref{pk}) for the power spectrum yields
(Taylor 1993)
\be
\label{pkz}
	P_\nonlin(\k) = \int d^3\!q \, e^{\im\k.\q}
	\{ e^{- k_i k_j [ \psi_{ij}(\bzero) - \psi_{ij}(\q) ]} - 1 \}
	\ .
\ee
This equation (\ref{pkz}), along with equations (\ref{psi}) 
and (\ref{psilin}), provides a nonlinear mapping between the 
initial and evolved 
power spectra, where the degree of evolution is controlled by the linear 
amplitude, $D(t)$, of perturbations.

The unit term on the right of the integrand of equation (\ref{pkz}),
which arises from subtraction of the mean density,
yields a delta-function at $\k = \bzero$ on integration.
The unit term can be ignored for $\k \neq 0$,
and we tacitly drop it hereafter.
Nevertheless, it can help in numerical integrations
to retain the unit term when $k$ is small (linear regime).

So far we have not assumed that the density field is isotropic,
although we have assumed it is homogeneous.
In Section \ref{Redsec} we will apply the power spectrum
equation (\ref{pkz})
in redshift space, where the density field is not isotropic.
In the remainder of the present Section
we assume that the density field is unredshifted and isotropic.

For an isotropic density field,
the displacement correlation function $\psi_{ij}(\q)$
resolves into irreducible components parallel and perpendicular
to the pair separation $\q$:
\be
\label{psipn}
	\psi_{ij}(\q)=\psi_{\p}(q) \hq_i \hq_j + 
	\psi_{\n}(q) (\delta_{ij} - \hq_i \hq_j)
	\ ,
\ee
where
\be
	\psi_{\p}(q) =
	{[D(t)]^2 \over 2 \pi^2} \int \dd k \, P_\lin(k)
	\left[ j_0(kq) - 2 \frac{j_1(kq)}{kq} \right]
\ee 
and 
\be
	\psi_{\n}(q) =
	{[D(t)]^2 \over 2 \pi^2} \int \dd k \, P_\lin(k)
	\left[ \frac{j_1(kq)}{kq} \right]
	\ .
\ee

For a general potential flow, as here,
the parallel variance $\psi_{\p}$ is related to the perpendicular variance
$\psi_{\n}$ by
(Monin \& Yaglom 1975; Gorski 1988)
\be
\label{irotcond}
	\psi_{\p}(q) = {\dd q \psi_{\n}(q) \over \dd q}
\ee
which can also be derived from equation (\ref{psilin})
with the definitions (\ref{psipn}).
It is convenient to define the differential displacement covariances
$\psi_+$ and $\psi_-$ by
\be
\label{psip}
	\psi_+(q) \equiv \psi_{\n}(\bzero) - \psi_{\n}(q)
\ee
\be
\label{psim}
	\psi_-(q) \equiv \psi_{\n}(q) - \psi_{\p}(q)
\ee
which are related by
\be
\label{psipm}
	\psi_-(q) = {q \, \dd \psi_+(q) \over \dd q}
	\ .
\ee
The (isotropic) power spectrum (\ref{pkz}) can then be written
in terms of $\psi_+$ and $\psi_-$
\be
\label{pkb}
	P_\nonlin(k) = \int \dd^3\!q \,
	e^{\im k q \mu - k^2 (\psi_+ + \psi_- \mu^2)}
\ee
where $\mu = \hat\k . \hat\q$.
Integrating equation (\ref{pkb})
over the azimuthal angle of $\q$ about the wavevector $\k$ is trivial,
and further integration over $\mu$ yields
\be
\label{pkc}
	P_\nonlin(k) = \Real \int_0^\infty \!
	2 \pi q^2\!\dd q \,
	e^{- k^2 (\psi_+ + \psi_-)}
	F(kq, k^2 \psi_-)
\ee
where $F(A,B)$ is the function
\be
\label{F}
	F(A,B) \equiv {\pi^{1/2} \over B^{1/2}}
	\exp \left( B - {A^2 \over 4 B} \right)
	\erf \left( B^{1/2} - {\im A \over 2 B^{1/2}} \right)
\ee
with $\erf(z)$ the error function
(Abramowitz \& Stegun 1968)
\be
	\erf(z) \equiv {2 \over \pi^{1/2}} \int_0^z \dd t \, e^{-t^2}
	\ .
\ee
The error function
$\erf(z)$ for complex $z$
is available in programs such as {\it Mathematica}.
Equation (\ref{pkc}) reduces the evaluation of the power spectrum
to a single integral over separation $q$.
Excepting some simple cases (Section \ref{power}),
this integral must be done numerically.

The integrand in equation (\ref{pkc})
oscillates rapidly when $kq$ is large and real,
which can make the integral difficult to evaluate
(Schneider \& Bartelmann 1995).
One way to resolve the difficulty is
to continue the $\psi_{\pm}(q)$ analytically into the complex plane, and
to shift the path of integration over $q$
into the upper complex plane,
which converts the oscillations into exponential decay.
Some experimentation may be required to determine a good
integration path, along which the integrand is well-behaved.
For example, in the case of an initially Gaussian power spectrum,
Section \ref{gauss} below,
we found it satisfactory to integrate first along the real axis
to $q \approx \min (2 D , 1/k)$,
and then to complete the integration along a straight line
slanted upwards in the complex plane,
at angle $\pi/6$ from the real line.
Variants on this path,
with or without the initial segment along the real axis,
and with other slant angles,
worked in other cases.

\begin{figure}
\epsfxsize=10.cm
\epsfysize=15.cm
\vspace{-3.cm}
\hfill
\epsfbox{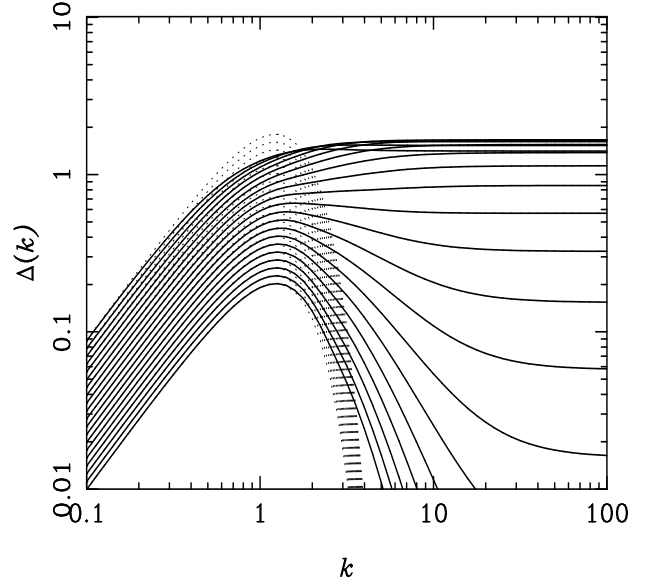}
\hfill
\epsfverbosetrue
\vspace{-4.cm}
  \small\caption{The evolution of the dimensionless amplitude
$\Delta_\lin(k)=k^{3/2} \exp -k^2/2$ in the Zel'dovich 
approximation for random Gaussian initial conditions. The solid
line is the nonlinear power spectrum and the dotted line is the
linear initial spectrum.}
\label{gaussfig}
\end{figure}
 
\subsection{Example: a Gaussian power spectrum}
\label{gauss}

	As an example, in Figure~\ref{gaussfig} we show the Zel'dovich
evolution of a power spectrum which is initially gaussian,
\be
	P_\lin(k) = \exp(-k^2)
	\ .
\ee
The quantity plotted is the dimensionless amplitude $\Delta(k)$,
the square root of the dimensionless power spectrum
defined by (Peacock 1992)
\begin{equation}
\label{Delta}
	\Delta^2(k) \equiv {k^3 P(k) \over 2 \pi^2}
	\ .
\end{equation}
Figure~\ref{gaussfig} shows clearly
the two main features of nonlinear growth,
the amplification of small scale power from the formation of caustics,
and the restriction on the growth of intermediate scale power
imposed by the positive density constraint in voids.

Figure \ref{gaussfig} also illustrates the limitations of the Zel'dovich
approximation at later stages.
While in reality the small scale density contrast increases
following the collapse of structures,
in the Zel'dovich approximation structures expand back outwards after collapse.
Thus in Figure~\ref{gaussfig}
the amplitude $\Delta(k)$ reaches a maximum of order unity
as a function of time, and thereafter declines.
In reality, the amplitude $\Delta(k)$ would be expected to
increase monotonically with time.


\subsection{Solutions for initial power--law spectra}
\label{power}

For initially power--law spectra, $P_\lin(k) \propto k^n$,
convergent results are attained in the Zel'dovich approximation
for power spectrum indices in the range $-3 < n < -1$.
For $n \leq -3$, the differential acceleration between particle pairs,
hence the differential displacement field,
receives divergent contributions from large scales,
while for $n \geq -1$, the acceleration diverges on small scales.
In the latter case, nonlinear processes are expected to intervene
so as to truncate the divergence,
but we do not consider such truncations here.

For initially power--law spectra
the differential displacement covariances
are power laws; $\psi_\pm \propto q^{-n-1}$.
It is convenient here
to normalize the growth factor, $D(t)$, so that
\be
\label{psipow}
	\psi_+(q) = D^2 q^{-n-1}
\ee
(a normalization based, for example, on the power spectrum is less convenient
because it leads to divergence in the displacement covariance for
$n=-3$ and $-1$).
The relation (\ref{psipm}) implies
\be
	\psi_-(q) = (-n\!-\!1) \psi_+(q)
	\ .
\ee
With the normalization (\ref{psipow}),
the initial linear power spectrum $P_\lin(k)$ is
(note $P_\lin(k)$ is defined without the factor $D^2$,
so $P(k) = D^2 P_\lin(k)$ in the linear regime)
\be
\label{pkpow}
	P_\lin(k) = 4\pi (2\!-\!n) \sin[(-n\!-\!1)\pi/2] \Gamma(1\!-\!n) k^n
\ee
which is positive over the interval $-3 < n < -1$,
going through zero at $n=-3$ and $-1$.

In the Zel'dovich approximation, initially power law power spectra
evolve in a self-similar fashion:
\be
\label{Self-sim}
	\Delta_\nonlin^2(k,D) = \Delta_\nonlin^2( D^2 k^{n+3} )
	\ .
\ee
The self-similar evolution of scale-free initial conditions
in the case of a flat, $\Omega = 1$, universe
is familiar from the BBGKY hierarchy (Peebles 1980).
In the Zel'dovich approximation,
the property of self-similarity extends also to non-critical Universes.

We now give exact solutions
for the cases $n=-1$, $-2$, and $-3$ in the Zel'dovich approximation.

\subsection*{\underline{(i) $n=-1$}}

In the limiting case $n \rightarrow -1$,
the displacement field has an incoherent Gaussian nature.
This arises because the incipient divergence of acceleration
on small scales causes particles at finite separations
to acquire essentially random velocities relative to each other.
But if particles in an initially uniform field move about randomly,
the resulting field continues to remain uniform.
Therefore, the power spectrum should remain infinitesimal even
when the displacement covariance grows to a finite value.
Mathematically, the linear power spectrum (\ref{pkpow}) goes to zero
in the limit $n \rightarrow -1$.

In reality, nonlinear evolution will in this case invalidate the linear
extrapolation of displacements assumed by the Zel'dovich approximation.
Nevertheless, it is interesting to consider $n=-1$ as a limiting case.
Defining the small positive quantity $\nu$ by
\be
	\nu \equiv -n\!-\!1
	\ ,
\ee
which tends to zero as $n \rightarrow -1$,
one finds that for $\psi_+ = D^2 q^\nu$, equation (\ref{psipow}),
the dimensionless power spectrum, $\Delta^2$, evolves as
\be
\label{n=-1}
	\Delta_\nonlin^2(k,D) = 3 \nu D^2 k^2 e^{- D^2 k^2}
	\ .
\ee
The power spectrum (\ref{n=-1}) 
is that of an $n=-1$ Gaussian field,
modulated by a random Gaussian distribution of displacements with
a one-dimensional one-point dispersion of $D$.

Although the power spectrum is itself infinitesimal,
the ratio $\Ps_2/\Ps_0$ of quadrupole--to--monopole redshift power,
which we shall consider in more detail later, 
is well-defined and finite in the limit $n \rightarrow -1$.

\subsection*{\underline{(ii) $n=-2$}}

For an initial power--law spectrum of index $n=-2$,
the integral (\ref{pkb}) can be done analytically.
Here $\psi_+ = \psi_- = D^2 q$,
and the dimensionless power spectrum evolves as
\be
\label{n=-2}
	\Delta_\nonlin^2(k,D) = {16 D^2 k \over \pi (1 + 4 D^4 k^2)^2}
	\left[ 1 + { 3\pi D^2 k \over 4 (1 + 4 D^4 k^2)^{1/2} } \right]
	\ .
\ee
In the linear regime, $\Delta_\nonlin^2 = 16 D^2 k/\pi$,
while in the highly nonlinear regime
$\Delta_\nonlin^2 \propto (D^2 k)^{-3}$.

\subsection*{\underline{(iii) $n=-3$}}

In the limiting case $n \rightarrow -3$,
differential accelerations between pairs are dominated by tidal contributions
generated on large scales,
inducing motions which appear locally like anisotropic Hubble flows.
The displacement covariances are
$\psi_+ = D^2 q^2$, $\psi_- = 2 \psi_+$.
The integral (\ref{pkb}) can again be done analytically,
yielding
\be
\label{n=-3}
	\Delta_\nonlin^2(k,D) = {e^{-1/(12 D^2)} \over (12\pi)^{1/2} D^3}
	\ .
\ee
Here the power spectrum remains a power--law, $P(k) \propto k^{-3}$,
at all times.
As argued by  Schneider \& Bartelmann (1995),
the persistence of the $k^{-3}$ form
results from the dominance of caustics in the density distribution.
The amplitude initially increases, from an exponentially tiny value,
as more regions reach the point of collapse,
but then reaches a maximum, at $D^2 = 1/18$,
and subsequently declines as $D^{-3}$
as regions pass through the point of collapse
and continue streaming outward.

We note that equation (\ref{n=-3}) also gives the asymptotic form of
the evolution of the power spectrum in the Zel'dovich approximation
on small scales whenever the spectral index is less than $-3$ on small scales.
This is true for example in the case of the Gaussian power spectrum
$P_L(k) = e^{-k^2}$ considered in Section \ref{gauss}.
In this case,
rms differential displacements between pairs are proportional
to their separation on small scales.
Define
\be
	\sigma^2 \equiv \lim_{q \rightarrow 0} {\psi_+(q) \over q^2} =
	{D^2 \over 60 \pi^2} \int_0^\infty \dd k \, k^2 P_\lin(k)
	\ .
\ee
Then for large $k$
(larger than the wavenumber at which the spectral index passes through $n=-3$),
\be
	\Delta_\nonlin^2(k,D) \approx
	{e^{-1/(12 \sigma^2)} \over (12\pi)^{1/2} \sigma^3}
\ee
which is the same as equation (\ref{n=-3}), with $\sigma$ in place of $D$.

\section{Redshift--space distortions}
\label{Redsec}

In Section \ref{Realsec} we considered some of the properties
of the density field evolving in the Zel'dovich approximation.
Here we turn our attention to the properties
of such a density field when viewed in redshift--space.
We assume the Zel'dovich approximation throughout this Section;
in Section 5 we will test the validity of the approximation with
$N$-body simulations.

\subsection{Zel'dovich approximation}

In redshift--space, the observed radial position of a galaxy is given by
its radial velocity,
which measures not only the uniform Hubble flow,
but also the peculiar motion of the galaxy.
Thus the position $\s$ of a galaxy in redshift space
appears shifted from its true position $\x$
by its radial peculiar velocity\footnote{ From hereon we
work in distance units of the velocity, where the Hubble parameter $H_0=1$.}:
\be
\label{eq:redmap}
	\s = \x + [\hxb.\vb(\x)] \hxb
	\ .
\ee
In the Zel'dovich approximation,
the peculiar velocity $\vb$ is related to the displacement $\xib$
at all times by the linear relation
$\vb[\x(\q),t] = \dot{\xib}(q,t) = \beta \xib(\q,t)$,
where $\beta$, equation (\ref{beta}),
is the dimensionless linear growth rate.
As discussed in the introduction, we assume an unbiased density field,
$b = 1$.
The position $\s$ of a galaxy in redshift space then appears
displaced from its initial Lagrangian position $\q$
(this initial position $\q$ is the same in both real and redshift space)
\be
\label{sq}
	\s = \q + \xibs(\q)
\ee
by a redshift displacement $\xibs(\q)$ which in the Zel'dovich approximation is
\be
\label{xibs}
	\xibs(\q)
	= \xib(\q) + \beta [\hxb.\xib(\q)] \hxb
	\ .
\ee
Equation (\ref{xibs}) shows that
the displacement $\xibs$ of a galaxy in redshift space
is stretched by a factor $(1+\beta)$ along the radial direction.
The overdensity $\delta^s$ in redshift space is then
(compare equation [\ref{e:6}])
\be
	\delta^s(\s) 	
	= \int \dd^3\!q \, \delta_D[\s-\q-\xibs(\q)] \ -\ 1
	\ .
\label{e:red}
\ee

The radial character of the redshift distortion
$\beta (\hxb.\xib) \hxb$
complicates taking the Fourier transform of the overdensity.
However, matters simplify
if the structure being observed is far away from the observer,
so that the distortions are effectively plane-parallel.
In the plane-parallel approximation,
the redshift displacement $\xibs$ becomes
\be
\label{xibsz}
	\xibs(\q)
	= \xib(\q) + \beta [\hz.\xib(\q)] \hz
\ee
where the line-of-sight direction $\hz$ is taken to be fixed.
The Fourier transform of the redshift density, equation (\ref{e:red}),
is then
\be
\label{deltask}
	\delta^s(\k) = \int \dd^3\!q \, e^{\im \k.\q} 
		(e^{\im \k.\xibs} -1)
	\ .
\ee
It is convenient to introduce the vector $\K$,
which is the wavevector $\k$ stretched by $(1+\beta)$ along the
$z$-axis:
\be
\label{K}
	\K = \k + \beta (\hz.\k) \hz
\ee
whose magnitude is
\be
\label{Kmag}
	K = k (1 + 2 \beta \mu_\k^2 + \beta^2 \mu_\k^2)^{1/2}
\ee
with $\mu_\k$ the cosine of the angle between the wavevector and
the line of sight
\be
	\mu_\k = \hz.\hat\k
	\ .
\ee
In terms of the vector $\K$,
equation (\ref{deltask}) can be rewritten
\be
\label{deltask2}
	\delta^s(\k) = \int \dd^3\!q \, e^{\im \k.\q} 
		(e^{\im \K.\xib} - 1)
	\ .
\ee

Just as the displacement $\xibs$ of a galaxy in redshift space
is stretched by a factor $(1+\beta)$ along the line-of-sight axis $\hz$,
equation (\ref{xibsz}),
so also is the displacement correlation function
$\psis_{ij} \equiv \lgl \xibs_i \xibs_j \rgl$
in redshift space stretched by $(1+\beta)$ factors along
each $z$-component of the tensor:
\be
\label{psis}
	\psis_{ij} = ( \delta_{ik} + \beta \hat{z}_i \hat{z}_k )
	\psi_{kl} ( \delta_{jl} + \beta \hat{z}_j \hat{z}_l )
	\ .
\ee
If the unredshifted displacement field $\xib$ is Gaussian
(as is true for an initially Gaussian field evolved
in the Zel'dovich approximation, Section \ref{Zeldsec}),
then the redshifted displacement field $\xibs$ is also Gaussian.

The derivation of expression (\ref{pkz}) for the evolved power spectrum
in Section \ref{Zeldsec}
did not assume isotropy, and remains valid in redshift space.
Thus the redshift power spectrum $\Ps(\k)$ is given by equation (\ref{pkz})
with the displacement correlation function taken to be the redshift
space quantity $\psis_{ij}$:
\be
\label{pksz}
	\Ps(\k) = \int d^3\!q \, e^{\im\k.\q}
	\{ e^{- k_i k_j [ \psis_{ij}(\bzero) - \psis_{ij}(\q) ]} - 1 \}
	\ .
\ee
As in equation (\ref{pkz}),
the unit term in the integrand can be dropped for $\k \neq 0$,
although again it helps in numerical integrations to retain the term when
$k$ is small.
Symmetry about the line of sight $\z$ implies that
the redshift power $\Ps(k,\mu_\k)$ is a function
of the magnitude $k$ of the wavevector $\k$
and of the cosine $\mu_\k$ of the angle between the wavevector and
the line of sight.

The relation (\ref{psis}) between the redshifted and unredshifted
displacement correlation functions $\psis_{ij}$ and $\psi_{ij}$ allows
the combination
$k_i k_j \psis_{ij}$
in the integrand of equation (\ref{pksz})
to be written conveniently in terms of the vector $\K$, equation (\ref{K}),
as $k_i k_j \psis_{ij} = K_i K_j \psi_{ij}$.
Then equation (\ref{pksz}) for the redshift power spectrum becomes
\be
\label{pks}
	\Ps(\k) = \int \dd^3\!q \,
	e^{\im\k.\q - K^2 (\psi_+ + \psi_- \mu^2)}
\ee
where now $\mu = \hat\K.\hat\q$,
and $\psi_\pm(q)$ are the (unredshifted) quantities given by equations
(\ref{psip}) and (\ref{psim}).
Equation (\ref{pks}) is the redshift space counterpart of equation (\ref{pkb}).
Integrating equation (\ref{pks})
over the azimuthal angle of $\q$ about the vector $\K$
yields a Bessel function $J_0$
\be
\label{pksa}
	\Ps(\k) = \int 2 \pi q^2\!\dd q \dd \mu \,
	e^{\im k q c \mu - K^2 (\psi_+ + \psi_- \mu^2)}
	J_0( k q s \sqrt{1-\mu^2} ) \nn
 \, 
\ee
where
$s$ and $c$ are
\be
\label{s}
	s = {\beta \mu_\k (1 - \mu_\k^2 )^{1/2} \over
	(1 + 2 \beta \mu_\k^2 + \beta^2 \mu_\k^2)^{1/2}}
	\ ,
\ee
\be
\label{c}
	c = (1 - s^2)^{1/2} = {1 + \beta \mu_\k^2 \over
	(1 + 2 \beta \mu_\k^2 + \beta^2 \mu_\k^2)^{1/2}}
	\ .
\ee

Except for the interesting case $s = 0$,
pursued further in Section \ref{approxP2P0},
equation (\ref{pksa}) cannot be reduced further to any simple analytic form.
Moreover, the oscillatory character of the integrand makes it difficult
to evaluate numerically.
Further details on the evaluation of the integral (\ref{pksa})
are given in Appendix A.
In Section \ref{smooth} we adopt a different approach to evaluating
the redshift power spectrum, which is to smooth it.

\subsection{Redshift power spectra for initial power--laws}

We now give analytic solutions for the evolution of the redshift power spectrum
in the Zel'dovich approximation for initially power-law spectra
$P_\lin \propto k^n$, for the cases $n=-1$ and $n=-3$.

In Section \ref{power} we
saw that the $n=-1$ spectrum gave rise to an incoherent 
random Gaussian velocity dispersion, resulting in a simple
damping term for the nonlinear power in the Zel'dovich approximation.
Extending this to the redshift domain, we can see that the 
effect will be an additional damping term in the line of sight
direction --- an approximation previously suggested by Peacock (1992)
to model the effects of virialised clusters.
With the normalisation discussed in Section \ref{power}, we find
that the redshift power spectrum is
\be
\label{red n=-1}
        \Delta_\nonlin^{\!s 2} (\k, D) =
        3 \nu D^2 k^2 (1 + \beta\mu_\k^2)^2 e^{- D^2 K^2}.
\ee
This is the same as the Peacock smoothing with
$\sigma^2_{\rm v} = D^2 (2 \beta + \beta^2)$, where $\sigma_{\rm v}$ is the
one-dimensional one-point velocity dispersion.
Notice that we also recover the linear Kaiser
boost factor, $(1 + \beta\mu_\k^2)^2$. Hence we conclude that
this form has some justification beyond
its phenomenological foundations for scales where the spectral index is
$\sim -1$, and arises as a special case in the model presented here.

For $n=-2$,
there appears to be no analytic solution for the redshift power spectrum,
unlike the unredshifted case, equation (\ref{n=-2}).

In the case $n=-3$, we find the analytic solution
\[
        \Delta_\nonlin^{\!s 2} (\k, D) =
        {1 \over (12\pi)^{1/2} D^3 (1 + 2\beta\mu_k^2 + \beta^2\mu_k^2)^{3/2}}
\]
\be
        \ \ \ \ \ \ \ \ \ \ \ \ \ \ \ \ \
        \exp \left[ -
        {1 + 2\beta\mu_k^2 + 3\beta^2\mu_k^2 - 2\beta^2\mu_k^4 \over
        12 D^2 (1 + 2\beta\mu_k^2 + \beta^2\mu_k^2)^2} \right]
\ee
which is the redshift counterpart of the unredshifted solution (\ref{n=-3}).

\subsection{Multipole analysis of the redshift power spectrum}

The effect of redshift distortion on the statistical
properties of the density field can be best interpreted
by a multipole expansion of the redshifted power spectrum. 
The anisotropic spectrum can be expanded in Legendre
polynomials
\be
	P^s(\k) = \sum_{\el=0}^{\infty} 
		P^s_{\el} (k)  \Pl (\mu_\k)
\ee
where $\Pl (\mu_\k)$ is the $\el^{th}$ Legendre polynomial and
$\mu_\k$ is again the cosine of the angle between the wavevector $\k$
and the line of sight.
The multipole moments of the power are given by
\be
\label{Psl}
	P^s_{\el} (k) = \frac{2 \el +1}{2} 
	\int_{-1}^{1} \dd \mu_\k \, P^s(\k) \Pl (\mu_\k).
\ee

Hamilton (1992) suggested using the ratios of the multipoles as a way of
measuring $\beta$, expanding the correlation function from 
Kaiser's linear theory. Cole, Fisher \& Weinberg (1994) 
developed this to study the anisotropy of the power spectrum,
where nonlinear effects could be avoided by limiting the analysis
to small wavenumbers. The distortion parameter,
$\beta$, could then be estimated from the ratio of monopole to quadrupole power:
\be
\label{e:Rlin}
	\frac{P^s_2(k)}{P^s_0(k)} =
	\frac{\frac{4}{3} \beta + \frac{4}{7} \beta^2}{
	1+ \frac{2}{3} \beta + \frac{1}{5} \beta^2}
	\ .
\ee
In the linear regime
each observable mode contributes to an independent estimate of $\beta$.

Nonlinear effects modify the quadrupole to monopole ratio
from the linear value (\ref{e:Rlin}).
In the following Section we derive an approximate expression
for this ratio in the Zel'dovich approximation,
and in subsequent Sections we present numerical results for the ratio
in the Zel'dovich approximation, in $N$-body simulations,
and in observations.

\subsection{An approximation to the quadrupole to monopole ratio}
\label{approxP2P0}

In this Section we derive an approximation, equation (\ref{pkse}),
which relates the ratio $\Ps_2/\Ps_0$
of the quadrupole to monopole redshift power spectrum
to the evolution of the unredshifted power spectrum
in the Zel'dovich approximation.

In equation (\ref{pksa}) for the redshift power spectrum,
notice that the Bessel function $J_0( k q s \sqrt{1-\mu^2} )$ 
in the integrand
equals one whenever $s$ is zero,
which happens
in the limit $\beta \rightarrow 0$ (equation [\ref{s}])
and also in the cases $\mu_\k = 0$ or 1
where the wavevector is perpendicular or parallel to the line of sight.
Moreover in the nonlinear limit where $k \rightarrow \infty$,
the $K^2$ terms in the exponential cause the integrand
of equation (\ref{pksa}) to decay rapidly as $q$ increases
and again the Bessel function is sensibly equal to one
(as is also the factor $e^{\im k q c \mu}$)
wherever the integrand is non-negligible.
In all these cases,
the redshift power, equation  (\ref{pksa}), reduces to the same form
as its unredshifted counterpart, equation (\ref{pkb}),
but with $\psi_\pm$ in the latter replaced by $(K/k)^2 \psi_\pm$,
where $K/k = 1+\beta\mu_\k^2$ (equation [\ref{Kmag}])
given that $s = 0$ and $c = 1$ (equation [\ref{c}]).
In other words, the redshift power equals the unredshifted power
evolved forward by $1+\beta\mu_\k^2$ in the growth factor $D(t)$:
\be
\label{pksb}
 	\Ps( \k , D ) \approx P [ k , (1 + \beta \mu_\k^2) D ]
	\ .
\ee
As it happens,
this same equation (\ref{pksb}) is also valid in the linear regime,
according to the usual linear relations
$P(k,D) = D^2 P_\lin(k)$ and
$\Ps(\k,D) = (1 + \beta \mu_k^2)^2 P(k,D)$.
Thus approximation (\ref{pksb}) is valid
(a) in the limit $\beta \rightarrow 0$,
or (b) if $\mu_\k = 0$,
i.e. if the wavevector $\k$ is perpendicular to the line of sight,
or (c) if $\mu_\k = 1$,
i.e. if the wavevector $\k$ is parallel to the line of sight,
or (d) in the highly nonlinear regime,
or (e) in the linear regime.
Evidently the approximation is of some generality.

%

Let us consider more closely the limiting case $\beta \rightarrow 0$,
for which the approximation (\ref{pksb}) becomes exact
(within the context of the Zel'dovich approximation).
Expanding equation (\ref{pksb}) as a Taylor series for small $\beta$ gives
\be
\label{pksc}
	\Ps(\k,D) = P(k,D)
	+ \beta \mu_k^2 D {\partial P(k,D) \over \partial D}
	\ \ \ (\beta \rightarrow 0)
	\ .
\ee
Here the redshift power spectrum is the sum of a constant term and
a term proportional to $\mu_k^2$,
which implies that the redshift power spectrum
is a sum of monopole and quadrupole terms, for small $\beta$.
The ratio of quadrupole to monopole power for small $\beta$ is,
from equation (\ref{pksc}),
\be
\label{pksd}
	{\Ps_2 (k) \over \Ps_0 (k)} =
	{4 \beta \over 3} \, {\partial \ln P(k,D) \over \partial \ln (D^2)}
	\ \ \ (\beta \rightarrow 0)
	\ .
\ee
Amongst other things,
equation (\ref{pksd}) predicts that the quadrupole redshift power $\Ps_2 (k)$
goes through zero where the unredshifted power $P[k,D(t)]$ reaches its
maximum as a function of time in the Zel'dovich approximation.

We now propose a generalisation of the result (\ref{pksd}) to arbitrary $\beta$,
which is consistent with and to some extent motivated by
approximation (\ref{pksb}),
and which accords with the numerical results of
Section \ref{Power law models} below,
which are further supported by the $N$-body simulations
in Section \ref{simulations}.
The numerical results indicate that,
at least in the case of initially power law power spectra,
the quadrupole to monopole ratio $\Ps_2/\Ps_0$
appears to satisfy a simple scaling law with $\beta$,
such the shape of the ratio is insensitive to $\beta$,
while its amplitude is proportional to the usual linear ratio
(strictly, the numerical results are for smoothed power spectra,
but we assume that the results hold also without smoothing).
We further find that the zero-crossing of the quadrupole
scales with $\beta$ such that $D^2 \sim 1/(1+\beta)$.
These empirical results, combined with equation (\ref{pksd}),
suggest the general approximation
\be
\label{pkse}
	{\Ps_2 (k) \over \Ps_0 (k)} \approx
	{\frac{4}{3} \beta + \frac{4}{7} \beta^2 \over
	1 + \frac{2}{3} \beta + \frac{1}{5} \beta^2} \,
	{\partial \ln P[k,(1+\beta)^{1/2}D] \over \partial \ln (D^2)}
	\ .
\ee
The approximation (\ref{pkse}) relates the quadrupole to monopole ratio
$\Ps_2/\Ps_0$
of the redshifted power spectrum to the evolution
$\partial \ln P / \partial \ln D^2$
of the unredshifted power spectrum in the Zel'dovich approximation.
Although we have numerical support for the approximation (\ref{pkse})
only in the case of initially power--law spectra,
we suggest that it is likely to be a good
approximation for arbitrary power spectra.

\subsection{The generating function of smoothed power spectrum multipoles}
\label{smooth}

So far we have obtained a number of analytic results
for the redshift power spectrum in the Zel'dovich approximation,
but in general it is necessary to resort to numerics.
Unfortunately,
numerical integration of the redshift power spectrum,
equation (\ref{pksa}), while feasible, is unpleasant (see Appendix A).
One way to sidestep the numerical difficulties is to smooth 
the power spectrum. 
This is not such a bad idea because it is necessary to
measure a smoothed power spectrum from observations.
That is, the Fourier modes measured in a catalogue are already
convolved with the window function of the catalogue,
and one may choose to smooth the power spectrum further to reduce error bars.
In addition, smoothing makes defining quantities such as the zero
crossing of the quadrupole, discussed in Section \ref{Power law models},
more robust to random errors.

The smoothed (unredshifted) power spectrum is defined as
\be
\label{Ps}
	\sP(\sk) = \int \dd^3\!k \, P(k) W_N(k,\sk)
	\ .
\ee
We choose to adopt a power law times Gaussian smoothing kernel
\be
\label{WN}
	W_N(k,\sk) = {k^{2N} \exp(-k^2/\sk^2) \over
	2 \pi \sk^{2N+3}\Gamma[(2N+3)/2)]}
	\ ,
\ee
which is analytically convenient,
and goes to a delta-function
in the limit $N \rightarrow \infty$. 
In subsequent Sections we use $N = 1$.

For the redshift power spectrum,
the smoothed harmonics can be defined similarly by
\be
\label{Pssl}
	\sPsl(\sk) = (2\el+1) \int \dd^3\!k \, \Ps(\k) \Pl(\mu_\k) W_N(k,\sk)
	\ .
\ee
The ratio $\sPs_0(\sk)/\sPs_2(\sk)$ of smoothed quadrupole--to--monopole 
harmonics
satisfies the usual equation (\ref{e:Rlin}) in the linear limit.

The whole hierarchy of smoothed spectra, for various $N$ and $\el$,
can be derived from the generating function
\be
\label{G}
	G ( a , b ) \equiv
	\int \dd^3\!k \, \Ps(\k)
	e^{- a k^2 - b k^2 \mu_\k^2}
	\ .
\ee
In terms of this generating function,
the monopole harmonic of the redshift power spectrum,
smoothed with the window $W_N$ is,
\be
\label{Pss0}
	\sPs_0(\sk) =
	\left[ - {\partial \over \partial a} \right]^{\!N}_{\!a=\sk^{-2}}
	\! 
\left[{G ( a , b=0 ) \over 2\pi\sk^{2N+3} \Gamma[(2N+3)/2]} \right]
\ee
while the quadrupole redshift power smoothed with the same window is
\[
	\sPs_2(\sk) = 
\]
\be
\label{Pss2}       
 \frac{5}{2}
\left[\frac{\partial}{\partial a} -3\frac{\partial}{\partial b}\right]_{b=0}
\left[ -\frac{\partial}{\partial a}\right]^{N-1}_{\!a=\sk^{-2}}
\!  \left[  { G ( a , b )\over 2\pi\sk^{2N+3} \Gamma[(2N+3)/2]} \right].
\ee

Substituting equation (\ref{pksz}) for the redshift power spectrum
into the generating function (\ref{G}) gives
\be
\label{Ge}
	G (  a , b ) =
	\int \dd^3\!q \, \dd^3\!k \,
	e^{i k . q - k_i k_j \Psi_{ij}(\q)}
\ee
where $\Psi_{ij}$ is the matrix
\be
\label{Psi}
	\Psi_{ij}(\q) = a \delta_{ij} + b \hat z_i \hat z_j
	+ \psis_{ij}(\bzero) - \psis_{ij}(\q)
	\ .
\ee
The $k$ part of the integral (\ref{Ge}) is the Fourier transform of a Gaussian,
which solves in the usual way to
\be
\label{Gq}
	G ( a , b ) =
	\int {\pi^{3/2} \dd^3\!q \over |\Psi|^{1/2}} \,
	e^{- q^2 \Psi_{qq}^{-1} / 4}
\ee
where $|\Psi|$ is the determinant of $\Psi_{ij}$,
\[
	|\Psi| = (\psi_+ + a)
	\{ (\psi_+ + \psi_- + a) [\psi_+ (1+\beta)^2 + a + b]
\]
\be
\label{detPsi}
	\ \ \ \ \ \ \ \ \ \ \ \ \ \ \ \ \ \ \ \ \ \ \ \ \ \ \ \ \ \ 
	+ \mu^2 \psi_- [a (2\beta+\beta^2) - b] \}
\ee
with $\mu = \hz.\hat\q$
and $\psi_\pm(q)$ given by equations (\ref{psip}) and (\ref{psim}),
and $\Psi_{qq}^{-1} = \hat q_i \Psi_{ij}^{-1} \hat q_j$ is the $qq$-component
of the inverse of $\Psi_{ij}$,
\[
	\Psi_{qq}^{-1} = {(\psi_+ + a) \over |\Psi|}
	\{ \psi_+ + a
\]
\be
\label{invPsi}
	\ \ \ \ \ \ \ \ \ \ 
	+ (1-\mu^2) [\psi_+ (2\beta+\beta^2) + b
	+ \mu^2 \psi_- \beta^2] \}
	\ .
\ee
Evaluation of the smoothed monopole and quadrupole spectra from
equations (\ref{Pss0}), (\ref{Pss2}) and (\ref{Gq})
still involves a double integral over $\mu$ and $q$.
However,
the integrand here is well-behaved (at least for small $N$, such as $N=1$)
and depends only on elementary functions (besides $\psi_\pm(q)$).
This compares to the unsmoothed case (Appendix A),
where the integrand is ill-behaved,
and evaluating the harmonics
involves a double integral over an infinite sum involving special functions.

\subsection{Numerical results for quadrupole-to-monopole power}
\label{Power law models}

\begin{figure}
\epsfxsize=10.cm
\epsfysize=15.cm
\vspace{-3.cm}
\hfill
\epsfbox{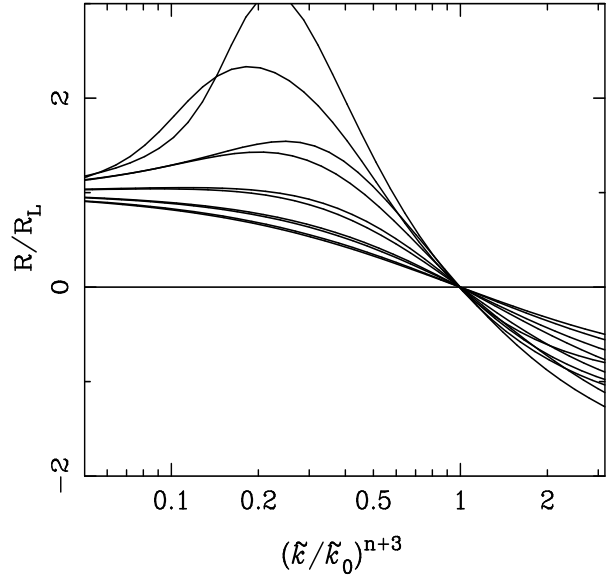}
\hfill
\epsfverbosetrue
\vspace{-4.cm}
\small\caption{Plots of the ratio $R$ of smoothed quadrupole-to-monopole
power against $(\sk/\sk_0)^{n+3}$
for initially power-law spectra with indices,
from lower curve to upper, of $n = -1$, $-1.5$, $-2$,$-2.5$ and $-2.9$.
In each case curves for $\beta = 0$ and 1 are shown,
which lie close to each other
(and practically on top of each other for $n=-1$ and $-1.5$)
when $R$ is scaled to the linear amplitude $R_\lin$ and the zero-crossing
scale $\sk_0$.
For each value of the spectral index,
$\beta = 0$ gives the slightly higher scaled ratio.
}
\label{R}
\end{figure}

From the generating function of multipole moments, equation (\ref{Gq}),
we have calculated numerically the smoothed quadrupole
and monopole moments of the power spectrum
for initially power law spectra $P_\lin \propto k^n$
with various values of the index $n$, and for various $\beta$.
The smoothing kernel is $W_N$, equation (\ref{WN}), with $N=1$.
Figure \ref{R} shows the ratio $R$ of the smoothed quadrupole-to-monopole power
\be
	R \equiv {\sPs_2(\sk) \over \sPs_0(\sk)}
\ee
for a representative sample of cases,
$n = -1$, $-1.5$, $-2$, $-2.5$ and $-2.9$,
and for $\beta = 0$ and 1 in each case.
Values intermediate between these give curves intermediate to those plotted.
The plotted ratio has been divided by the linear ratio,
$R_\lin = (\frac{4}{3} \beta + \frac{4}{7} \beta^2)/
( 1 + \frac{2}{3} \beta + \frac{1}{5} \beta^2)$,
equation (\ref{e:Rlin}),
and the wavenumber has been scaled to the zero-crossing scale, $\sk_0$.

It is immediately apparent from Figure \ref{R}
that the shape of the quadrupole-to-monopole ratio $R$
depends on the spectral index $n$, but is insensitive to $\beta$
at fixed $n$, at least for $n \ga -2.5$.
The insensitivity to $\beta$ is an interesting result,
and we have previously invoked this insensitivity in proposing the
approximation (\ref{pkse}) to the (unsmoothed) ratio.
As to the spectral index,
Figure \ref{R} shows that for $n \la -2$,
the ratio $R$ actually rises up above its linear value
before turning over and going through zero.
Physically,
the linear squashing effect is first enhanced by caustics
in redshift space as structures approach turnaround,
and is then subsequently negated by ``fingers-of-god''
as structures collapse.
Figure \ref{R} indicates that for $n \la -2$
the caustic effect wins over the finger-of-god effect at translinear scales.
However, as the enhancement peak scales as $\sk_{\rm pk}\sim
0.2^{1/(n+3)}\sk_0$, we find that $R$ decreases for  $n \la -2$,
for fixed $\beta$, in the regime $k\sim k_0$.

A useful fitting function to the curves in Figure \ref{R} is
\be
\label{e:fitfunc}
	R = R_\lin e^{-(n+1)/2}[1-(\sk/\sk_0)^{0.6(n+3)}]\ ,
\ee
which is accurate to within $10$ percent
over the range $0.1 < \sk/\sk_0 < 1$, and $-3 < n < -1$. 
We shall use the fitting function (\ref{e:fitfunc})
in fitting the observed ratio $R$ measured in the QDOT plus 1.2 Jy survey
to $\beta$ and $n$,
Section \ref{observations}.

The zero-crossing scale of the quadrupole power also contains useful
information about $\beta$ and the spectral index.
The zero-crossing scale discussed below refers strictly to that of the smoothed
quadrupole power, but the zero-crossing scale of the unsmoothed quadrupole
is not greatly different.

Suppose that the zero-crossing of the smoothed quadrupole power
$\sPs_2(\sk)$ is observed to occur at wavevector $\sk_0$.
Define a corresponding real scale by $\tq_0 \equiv \alpha/\sk_0$,
where $\alpha$ is a fitting constant, discussed below.
Then we find that
the dimensionless amplitude of the displacement correlation function
at the zero-crossing scale $\tq_0$
is fitted tolerably well by the following fitting formula
\be
\label{zerox}
	{\psi_+(\tq_0) \over \tq_0^2} \approx {C \over 1+\beta}
	\ ,\ \ \ 
	\tq_0 = \alpha/\sk_0
	\ .
\ee
In the Zel'dovich approximation,
formula (\ref{zerox}) with the values
(we adopt here the value of $\alpha$ which best fits the $N$-body simulations
reported in Section \ref{simulations},
since the same value works well also for Zel'dovich)
\be
\label{fitpar}
	\alpha = 2.3
	\ , \ \ \ 
	C = 0.19
\ee
is accurate to better than 20\% for initially power-law spectra
over the observationally interesting range of spectral indexes
$-2 \le n \le -1$
and linear growth rate parameter
$0 \le \beta \le 1$.
As a matter of interest,
in the case $n=-1$ the exact analytic result (in the Zel'dovich approximation)
is
$\psi_+(\tq_0) / \tq_0^2 = 1/[\alpha^2(1+\beta)]$.
Physically,
it is not too surprising that the zero-crossing of the quadrupole should occur
when the rms displacement $\psi_+^{1/2}(q)$ between pairs has reached an
appreciable fraction of their separation $q$.

The amplitude of the power spectrum itself at the zero-crossing $\sk_0$
is a more complicated function of spectral index.
During linear evolution,
the dimensionless smoothed monopole redshift power
$\tilde\Delta^{\!s 2}_0(\sk)$
is related to the dimensionless displacement correlation function
$\psi_+(\tq)/\tq^2$ at $\tq = \alpha/\sk$
by (compare equation [\ref{pkpow}];
note the factor $\Gamma[(5\!+\!n)/2] / \Gamma(5/2)$ below comes from smoothing)
\[
	\tilde\Delta^{\!s 2}_0(\sk) =
	(1\!+\!\frac{2}{3}\beta\!+\!\frac{1}{5}\beta^2)
	{2 \over \pi} (2\!-\!n) \sin[(-n\!-\!1)\pi/2] \Gamma(1\!-\!n)
\]
\be
\label{zeroxP}
\ \ \ \ \ \ \ \ \ \ \ \ 
	{\Gamma[(5\!+\!n)/2] \over \Gamma(5/2)}
	\alpha^{3+n} {\psi_+(\alpha/\sk) \over (\alpha/\sk)^2}
	\ .
\ee
Nonlinear evolution, which is starting to become important at zero-crossing,
will modify the power from the linearly evolved value (\ref{zeroxP}).
Setting aside this effect of nonlinearity,
one concludes from equation (\ref{zeroxP}), combined with the approximation
(\ref{zerox}), that the amplitude of the monopole power at zero-crossing
is sensitive to the spectral index $n$, but relatively insensitive to $\beta$
(the $1/(1+\beta)$ factor in equation [\ref{zerox}] tending to cancel
the $1+\frac{2}{3}\beta+\frac{1}{5}\beta^2$ factor in equation [\ref{zeroxP}]).

It is worth emphasising this interesting if somewhat disappointing result:
one might have hoped that the amplitude of the observed monopole power at the
zero-crossing of the quadrupole would provide a measure of $\beta$ in
the mildly nonlinear regime, offering a check on the value of $\beta$
measured from the quadrupole-to-monopole ratio in the linear regime,
and perhaps a test of the bias $b$.
However, the hope fails:
the power at zero-crossing is sensitive mainly to the spectral index,
not to $\beta$.

\section{Comparison with Simulations and Observations}
\label{Obssec}

\subsection{Comparison with simulations}
\label{simulations}

\begin{figure}
\epsfxsize=10.cm
\epsfysize=15.cm
\vspace{-3.cm}
\hfill
\epsfbox{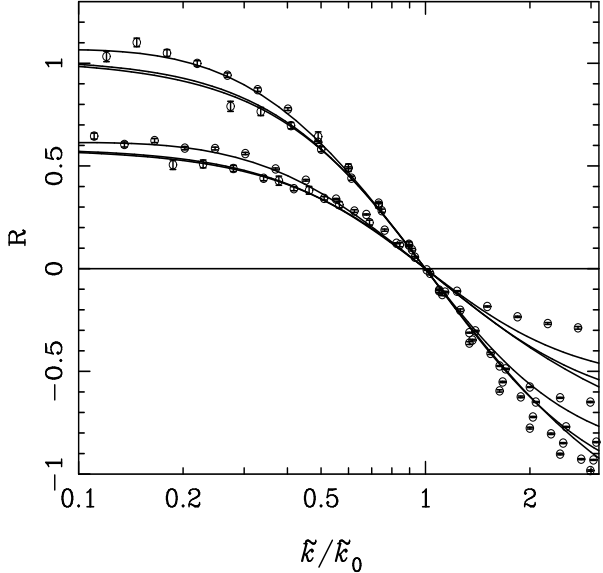}
\hfill
\epsfverbosetrue
\vspace{-4.cm}
\small\caption{ Plots of the ratio $R$ of smoothed monopole--to--quadrupole
power scaled by the zero-crossing, $\sk_0$. 
From top to bottom on scales $\sk<\sk_0$ are plotted the 
numerical results for the Zel'dovich evolution
for each power spectrum (solid lines) for 
$\beta=1$ with $n=-1$, $-1.5$ and $-2$ and $\beta=0.5$
with $n=-1$, $-1.5$ and $-2$. Overlayed are the results 
from a set of N-body simulations with the same initial parameters (points).
 Errors on the simulations are from cosmic variance.
}
\label{nbodfig}
\end{figure}

Given the approximate nature of the Zel'dovich mapping
assumed in Sections \ref{Realsec} and \ref{Redsec},
it is important to check the accuracy of 
these results with $N$-body simulations, where nonlinear
effects can be accounted for on smaller scales. The simulations 
we ran used the AP$^3$M code of Couchman (1991), generalised to
low-$\Omega$ by Peacock \& Dodds (1995), on a $64^3$ grid 
with $64^3$ particles.

We ran simulations with power law initial spectra with indices
$n=-1$, $-1.5$ and $-2$,
both for $\Omega=1$,
and for a low-$\Omega$ spatially flat model with 
$\Omega=0.3$ and $\Omega_{\Lambda}=0.7$, corresponding to $\beta = 0.5$.
In each case three simulations were computed
with differing random Gaussian initial conditions,
18 simulations altogether.
We assumed no bias in the density
field, and evolved all the models for 300 time steps so as
to erase the effects of using the Zel'dovich approximation
in the initial setting up of the simulation.

Power multipoles were calculated
by assigning densities to the grid by the cloud-in-cell algorithm and
FFT-ing the cube.  The shot noise component of the power was 
subtracted and the resulting Fourier space compressed by
averaging over azimuthal angles. The multipoles were then calculated 
by a least squares fit to each mode:
\be
	\chi^2(k) = \sum_i w_i(k) \left[\tilde{\Delta}^2(k,\mu_i) - \sum_{\el}
		 \tilde{\Delta}^2_{\el} (k)  \Pl (\mu_i) \right]^2.
\ee
The weighting scheme used was the inverse variance, 
$w_i(k)=1/\sigma^2(k,\mu_i)$, calculated from the intrinsic cosmic variance, 
$\sigma(\k)=\sqrt{2} \Delta^2(\k)$, for each mode. 
The resulting power multipoles were then smoothed using the kernel function
$W_N$ with $N=1$, equation (\ref{WN}),
and an inverse variance weighting scheme.

Generally, we found that the Zel'dovich approximation ceased to provide
a good approximation to the evolution of monopole or quadrupole power
at moderately nonlinear epochs.
This is to be expected, since as discussed in Section \ref{gauss},
the expansion of structures following collapse which occurs in the
Zel'dovich approximation causes power to reach a maximum and then decline,
whereas in reality power continues to increase monotonically.

However,
as shown in Figure \ref{nbodfig},
the Zel'dovich approximation does appear to reproduce remarkably well
the amplitude and shape of the ratio $R$ of the quadrupole-to-monopole power
seen in the $N$-body simulations,
for scales down to the zero-crossing of the quadrupole,
$\sk \leq \sk_0$.
Small deviations occur at small $\sk$, but these can be attributed to
the lack of contribution from smaller wavenumbers
in the smoothing process of the simulations. 
Small deviations are also seen in the $n=-2$ spectra,
at least some of which may arise from cosmic variance and the
influence of the finite box size on the longest wavelengths
in the simulations.
But the change in shape of the quadrupole--monopole ratio with spectral index
predicted by the Zel'dovich approximation is clearly seen in the simulations.
All the models fail in the regime $\sk > \sk_0$,
where we expect virialised clusters to provide the strongest 
redshift distortions (Jackson 1972).

The scaling relation suggested by equation (\ref{zerox})
for the amplitude of the displacement covariance at the zero-crossing
of the quadrupole is also found to hold for the simulations.
However, the amplitude $C$ predicted by the Zel'dovich 
approximation was not so accurate.  Applying a least squares 
fit of the model to the ensemble of 18 simulations, we found
\be
\label{alphaC}
	\alpha = 2.3 \pm 0.4
	\ ,\ \ \ 
	C = 0.54 \pm 0.07
\ee
with a formal $\chi^2=15.6$ for 16 degrees of freedom.
In general while the form of the scaling relation (\ref{zerox}) holds,
the Zel'dovich approximation underestimates the wavenumber at zero-crossing
of the quadrupole by a factor of approximately $2$ in the cases considered.
This can be understood as arising from the outflow of particles which follows
caustic formation in the Zel'dovich approximation.
This outflow exaggerates the ``finger-of-god'' effect,
causing the zero-crossing of the quadrupole to appear at larger scales
in the Zel'dovich approximation than is actually the case.

In fitting to observed data in the next Section,
we use the $N$-body simulations 
to calibrate the zero-crossing of the quadrupole.

\subsection{Application to the QDOT plus 1.2 Jy redshift survey}
\label{observations}

\begin{figure}
\epsfxsize=10.cm
\epsfysize=15.cm
\vspace{-3.cm}
\hfill
\epsfbox{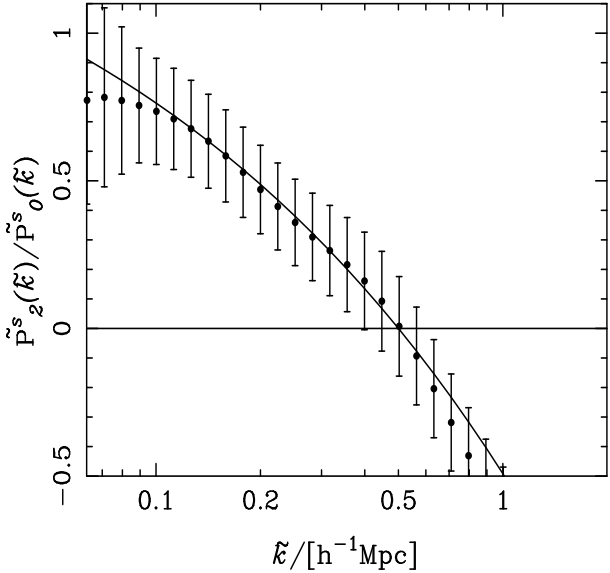}
\hfill
\epsfverbosetrue
\vspace{-4.cm}
  \small\caption{Plot of the ratio $R$ of smoothed monopole-to-quadrupole
power as a function of wavenumber $k$ for the merged QDOT plus 1.2 Jy redshift
survey.  The solid line is a fit to a model with
$\beta=0.8$ and $n=-2.5$, and with $\sk_0=0.5 h{\rm Mpc}^{-1}$.
\label{QDOT-1.2}}
\end{figure}

We have applied the results obtained above to estimate the distortion parameter,
$\beta$, and the local primordial spectral index, $n$,
from the amplitude and shape of the ratio $R$ of smoothed quadrupole-to-monopole
power observed in the merged QDOT plus 1.2 Jy {\it IRAS} redshift survey.
We further infer the
variance $\sigma_8^2$ of counts in $8\, h^{-1} {\rm Mpc}$ spheres
from the observed scale of the zero-crossing of the quadrupole.
The observed quadrupole-to-monopole ratio, taken from Hamilton (1995),
is shown in Figure \ref{QDOT-1.2}.
Superimposed on the observations is
a model with $\beta =0.8$ and $n=-2.5$.
The zero-crossing is estimated to be at $\sk_0=0.5 \pm 0.1 h{\rm Mpc}^{-1}$.

We use the fitting function (\ref{e:fitfunc})
to fit the amplitude, which depends on $\beta$, and the shape,
which depends on $n$, of the observed quadrupole-to-monopole ratio $R$.
We construct a likelihood function 
\be
	{\cal L} = \exp - a_i M_{ij} a_j
\ee
where $a_i = (\rm{data} - \rm{model})$ is the difference between
the $i^{th}$ data  point and the model prediction, and $M_{ij}$
is the data covariance matrix whose diagonal is the uncertainty 
on each data point and whose off-diagonal terms are estimated
assuming the underlying data have independent Gaussian-distributed
errors, and the major source of correlations comes from smoothing.
As our model breaks down on scales smaller than the 
zero--crossing point, and the errors on the observed quadrupole 
diverge at low wavenumber, we restrict our analysis to 
wavemodes in the range $0.06 < \sk/(h{\rm Mpc}^{-1})<0.5$.

\begin{figure}
\epsfxsize=10.cm
\epsfysize=15.cm
\vspace{-3.cm}
\hfill
\epsfbox{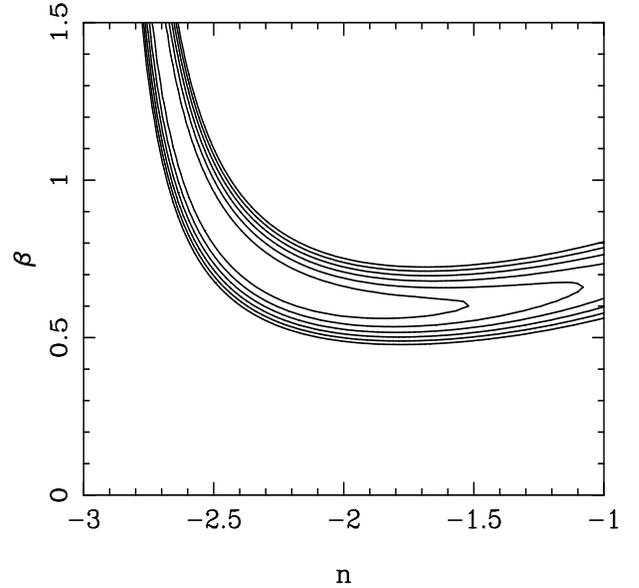}
\hfill
\epsfverbosetrue
\vspace{-4.cm}
  \small\caption{Likelihood function ${\cal L}(\beta,n|{\rm data})$,
of $\beta$ and the local spectral index, $n$, from the quadrupole-to-monopole 
ratio of the merged QDOT plus 1.2 Jy redshift survey. Contours are spaced by 
$\delta \ln {\cal L}=-0.5$ with the inner contour delineating 
the $68 \%$ 
and outer contour the $95 \%$ confidence region.  While the
spectral index is 
not strongly constrained, $\beta<0.5$ is ruled out at the $95\%$
level. This rules out $\Omega<0.3$ if
{\it IRAS} galaxies are unbiased.
\label{beta_v_n} }
\end{figure}

Figure \ref{beta_v_n} shows the marginalised likelihood 
in the $(\beta,n)$ plane. As $n \rightarrow -3$, $\beta$ is 
forced to infinity by our fitting function. 
The likelihood function selects this 
range as the better fit to the data due to the slight inflection
in the quadrupole--monopole ratio around $\sk=0.2 h \rm{Mpc}^{-1}$.
However if we consider only the $95 \%$ confidence limit 
given by the outer contour, low values of $\beta<0.5$ are
strongly excluded. In terms of the galaxy bias parameter, this
implies that the density parameter $\Omega > 0.3 b^{5/3}$. The 
value of the primordial spectral index on these scales is not
strongly constrained, but is consistent with $n \approx -2$.

\begin{figure}
\epsfxsize=10.cm
\epsfysize=15.cm
\vspace{-3.cm}
\hfill
\epsfbox{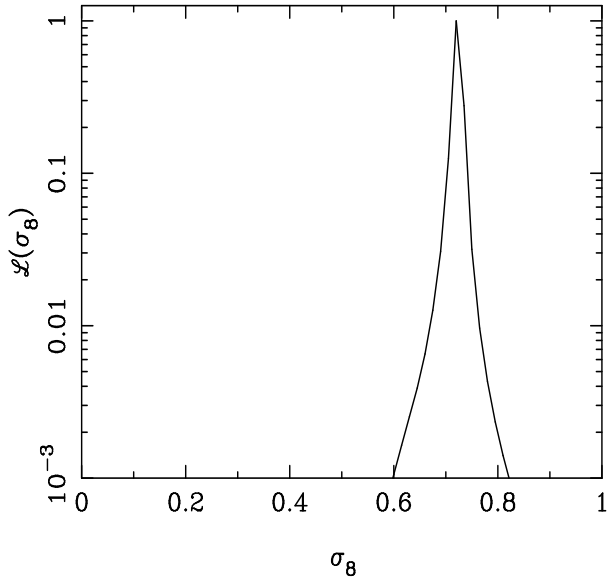}
\hfill
\epsfverbosetrue
\vspace{-4.cm}
  \small\caption{Marginalised likelihood function ${\cal L}(\sigma_8)$
of the variance in $8\,h^{-1}{\rm Mpc}$ spheres inferred
from the zero-crossing of the QDOT plus 1.2 Jy quadrupole.
\label{like_sig8}
}
\end{figure}

We estimate the clustering amplitude $\sigma_8$
from the scale of the zero-crossing of the quadrupole,
using equation (\ref{zerox}) with the parameters (\ref{alphaC})
calibrated from the $N$-body simulations.
In figure \ref{like_sig8} we plot the marginalised likelihood function 
${\cal L}(\sigma_8) =
\int\!\dd \beta \dd n {\cal L}(\beta,n,\sigma_8|{\rm data})$.
We deduce that 
$\sigma_8=0.7 \pm 0.05$
where we quote $95 \%$ errors. This is in very good agreement
with values found by other methods for the {\it IRAS} galaxy
catalogs (eg, Fisher et al. 1994, Heavens \& Taylor 1995) if
IRAS galaxies are unbiased. Given the indirect method of our
measurement, we find this very encouraging.

\section{Summary}
\label{Sumsec}
	
The main aim of this paper has been to study the redshift distortion
of the power spectrum in the moderately nonlinear regime.
Our approach is 
motivated by the consideration that structures may appear in redshift space
more nonlinear than they really are ---
for example, regions which are just turning around in real space
appear as caustics in redshift space.
This leads us firstly to work in Lagrangian space,
and secondly to use the Zel'dovich approximation,
which is in effect linear theory expressed in Lagrangian space.

We started by deriving an expression relating the power spectrum
to the Lagrangian displacement field.
We used this to determine the evolution of the power spectrum
in the Zel'dovich approximation, first in real space,
then in redshift space in the plane-parallel approximation.
We presented some analytic solutions for initially power-law spectra.
In particular, we showed that a spectrum with index $n=-1$
gives rise to an incoherent Gaussian displacement field,
producing nonlinear redshift distortions with the same form as
the Kaiser-Peacock model.

We derived various analytic and numerical results for the observationally
interesting ratio $R$ of quadrupole-to-monopole redshift power,
whose value in the linear regime,
$R_\lin = (\frac{4}{3}\beta+\frac{4}{7}\beta^2)/
	(1+\frac{2}{3}\beta+\frac{1}{5}\beta^2)$,
yields a measure of the distortion parameter $\beta$.
In the Zel'dovich approximation,
the amplitude of the ratio $R$ is set by the linear value $R_\lin$,
but the shape of $R$ as a function of wavenumber $k$
depends mainly on the spectral index $n$,
and is insensitive to $\beta$ at fixed $n$.
The zero-crossing of the quadrupole power
occurs at the point where the Zel'dovich power spectrum
is a maximum as a function of time.

We have tested the Zel'dovich results against $N$-body simulations
with initially power law spectra, in both $\Omega=1$ and low-$\Omega$ models.
The simulations show that the Zel'dovich approximation
ceases to provide a good approximation to the power spectrum
at moderately nonlinear epochs.
Remarkably however,
the Zel'dovich approximation predicts rather well the
amplitude and shape of the quadrupole-to-monopole ratio $R$
on scales down to the zero-crossing of the quadrupole,
when $R$ is scaled to the scale of the zero-crossing.
The Zel'dovich approximation underestimates the wavenumber at zero-crossing
by a factor of about two,
although it predicts correctly the way the zero-crossing scales with
$\beta$ and $n$.

We have applied these findings to
estimate the distortion parameter $\beta$,
the local spectral index $n$,
and the variance $\sigma_8^2$ of counts in $8\, h^{-1} {\rm Mpc}$ spheres,
from the quadrupole-to-monopole ratio $R$
measured in the merged QDOT plus 1.2 Jy redshift survey.
We find that the distortion parameter is constrained to $\beta>0.5$
at the 95\% level.
The spectral index is not well constrained,
but is consistent with $n\approx -2$ at translinear scales.
The clustering amplitude $\sigma_8$, inferred from the scale of the
zero-crossing of $R$,
is $\sigma_8=0.7\pm0.05$,
consistent with other estimates. 

The success of the Zel'dovich model
in describing the quadrupole-to-monopole ratio $R$
suggests that departures from the linear value $R_\lin$
at translinear scales are caused mainly by infall on to clusters,
not by virialised cluster cores.

\bigskip
\noindent{\bf ACKNOWLEDGMENTS}
\bib\strut

\noindent
ANT is supported by a PPARC research assistantship, and
thanks Alan Heavens and John Peacock for stimulating 
discussion.
AJSH thanks George Efstathiou for the hospitality of the
Nuclear and Astrophysics Laboratory at Oxford University,
where AJSH enjoyed the support of a PPARC Visiting Fellowship
during 1994/5,
and he thanks John Peacock for hospitality during a visit to ROE,
where this collaboration began.
AJSH appreciates support from NSF grant AST93-19977,
and NASA Astrophysical Theory Grant NAG 5-2797.

While in the process of completing this work
we became aware of a similar paper by Fisher \& Nusser (1995)
who reach similar conclusions.

\bigskip
\noindent{\bf REFERENCES}
\bib \strut

\bib Abramowitz M. and Stegun I.A., 1968, Handbook of 
Mathematical Functions, Dover, New York

\bib Bond J.R., Couchman H.M.P., 1987, preprint

\bib Ballinger W. E., Heavens A. F., Taylor A. N., 1995, MNRAS, 276, L59

\bib Cole S., Fisher K.B., Weinberg D.H., 1994, MNRAS, 267, 785

\bib Couchman H.M.P., 1991, ApJ, 368, L23

\bib Efstathiou G., 1990, in Peacock J.A., Heavens A.F., Davies A., 
eds, Proc. 36th  Scottish Universities Summer School in Physics,
{\em Physics of the Early Universe}, Adam Hilger, Bristol

\bib Fisher K.B., Scharf C.A., Lahav O., 1994, MNRAS, 266, 219

\bib Fisher K.B., Lahav O., Hoffman Y., Lynden--Bell D., Zaroubi S., 
1995, MNRAS, 272, 885

\bib Fisher K.B., Nusser A., 1995, preprint

\bib Gorski K.M., 1988, ApJ Lett, 332, L7

\bib Hamilton A.J.S., 1992, ApJ, 385, L5

\bib Hamilton A.J.S., 1995, in {\em Clustering in the Universe},
Proc. XVth Rencontres de Moriond, in press.

\bib Hamilton A.J.S., Culhane M., 1996, MNRAS, 278, 73

\bib Heavens A., Taylor A.N., 1995, MNRAS, 275, 483

\bib Hivon E., Bouchet F.R., Colombi S., Juszkiewicz R., 1995,
A.A., 298, 643

\bib Hui L., Bertschinger E., 1995, preprint

\bib Jackson J.C, 1972, MNRAS, 156, 1p

\bib Kaiser N., 1987, MNRAS, 227, 1.

\bib Mann R.G., Heavens A.F., Peacock J.A., 1993, MNRAS,
263, 798

\bib Monin A.S., Yaglom A.M., 1975, {\em Statistical Fluid
Mechanics}, Vols. 1 \& 2, MIT Press, Cambridge.

\bib Peacock J.A., 1992 in Martinez V., Portilla M., S\'{a}ez D., eds, 
{\em New Insights into the Universe}, Proc. Valencia summer school. 
Springer,Berlin, p.1

\bib Peacock J.A., Dodds S.J., 1994, MNRAS, 267, 1020

\bib Peebles P.J.E., 1980, {\sl Large Scale Structure in the Universe},
Princeton University Press

\bib Schneider P., Bartelmann, M, 1995, MNRAS, 273, 475

\bib Taylor A.N., 1993, in Proc. Cosmic Velocity Fields, 9$^{th}$ IAU
Conf, eds F. Bouchet M. Lachieze-Rey, Edition Fronti\`{e}re, Gif-sur-Yvett,
p585

\bib Zaroubi S., Hoffman Y., preprint 1994

\bib Zel'dovich Y.B., 1970, A\&A, 5, 84.

\section*{APPENDIX A}
\label{Apk}

This Section shows how to evaluate the `difficult' integral (\ref{pksa})
for the redshift power spectrum in the Zel'dovich approximation.
The integral is a two-dimensional integral over $\mu = \hat\K . \hat\q$
and separation $q$.
We choose to develop the integral over $\mu$ analytically
as an infinite sum,
both because the numerical integral over $\mu$ is itself unpleasant,
and because the expression as an infinite sum allows the $q$-integrand
to be continued analytically into the complex $q$-plane,
which stabilizes the integration over $q$.

For small $k q s$,
the Bessel function $J_0( k q s \sqrt{1-\mu^2} )$ in the integrand
can be expanded as a power series in $1-\mu^2$.
Integration of the leading term (which is 1) in the series leads
to an expression similar to (\ref{pkc}),
and higher terms in $1-\mu^2$ can be generated by repeated differentiation
of this expression.
The result is the infinite sum
\[
	\Ps(\k) = \Real \int_0^\infty 2\pi q^2\!\dd q \,
	e^{- K^2 (\psi_+ + \psi_-)}
\]
\be
\label{Apk1}
	\ \ \ \ \ \ \ \ \ \ 
	\sum_{n=0}^\infty {(-)^n (k q s / 2)^{2n} \over n!^2}
	\left( {\partial \over \partial B} \right)^{\!n}_{\! B = K^2 \psi_-}
	\! F(k q c , B)
\ee
where $F$ is the function defined by equation (\ref{F}),
and $s$ and $c$ are given by equations (\ref{s}) and (\ref{c}).
The sum in equation (\ref{Apk1}) can be truncated to a few terms if
$k q s$ is small.
For large $k$ (nonlinear limit),
this truncation effectively works for all $q$,
since then the exponential factor $e^{-K^2 (\psi_+ + \psi_-)}$
causes the integrand to become negligible when $k q s$ is not small.

The expansion (\ref{Apk1}) fails when $k q s$ is large
and at the same time $K^2 (\psi_+ + \psi_-)$ is small.
Here a series expansion in $K^2 \psi_-$
leads to the following infinite sum:
\[
	\Ps(\k) = \Real \int_0^\infty 4\pi q^2\!\dd q \,
	e^{- K^2 \psi_+}
\]
\be
\label{Apk2}
	\ \ \ \ \ \ \ \ \ \ 
	\sum_{n=0}^\infty {(K^2 \psi_-)^n \over n!}
	\left( {\partial \over \partial A} \right)^{\!2n}_{\! A = k q c}
	{e^{\im [A^2 + (k q s)^2]^{1/2}} \over \im [A^2 + (k q s)^2]^{1/2}}
	\ .
\ee
Expression (\ref{Apk2}) works when $K^2 \psi_-$ is small and
$k q s$ is not small,
which is complementary to the domain of validity of equation (\ref{Apk1}).
A program such as {\it Mathematica} helps considerably
with the algebra involved in taking the repeated derivatives in both
equations (\ref{Apk1}) and (\ref{Apk2}).

The integration of equations (\ref{Apk1}) and (\ref{Apk2}) over $q$
must be done numerically.
As in the unredshifted case,
the oscillatory character of the integrand for large $k q$
can be converted into exponential decay by shifting the path of integration
of $q$ into the upper complex plane.
One should be careful here that
although the real parts of the integrands of
equations (\ref{Apk1}) and (\ref{Apk2})
agree for real $q$,
their imaginary parts differ, so that their analytic continuations in the
complex $q$-plane differ.
The integrands could be made identical by adding their
complex conjugate expressions,
but this would introduce an undesirable exponentially increasing component
in the upper complex plane.
Therefore, if it is necessary to use both expressions
(\ref{Apk1}) and (\ref{Apk2}) in integrating from $q = 0$ to $\infty$,
then the transition from one expression to the other
along the path of integration must occur on the real line.

In the cases we tried,
an integration path similar to that described
at the end of Section \ref{Zeldsec} proved satisfactory.
Namely, integrate first some way along the real axis,
then complete the integration along a straight line upwardly slanted
in the complex plane.
In the case $K^2 \psi_-(q = 1 / k s) \geq 1$,
equation (\ref{Apk1}) works everywhere on the path of integration.
In the case
$K^2 \psi_-(q = 1 / k s) < 1$,
equation (\ref{Apk1}) works for the first part of the integration,
along the real line,
but equation (\ref{Apk2}) is required for the second part,
a suitable path being vertically upwards in the complex plane.

Computing the quadrupole and monopole harmonics of the redshift power spectrum
requires a final numerical integration of $\Ps(\k)$
over $\mu_\k$ with $\Pl(\mu_\k)$ for $\el = 0$ and 2,
equation (\ref{Psl}).

\end{document}